\documentclass[preprint,prb,aps,showkeys]{revtex4}
\usepackage{graphicx}
\usepackage{enumitem}   
\usepackage{amsmath,amssymb}

\begin{document}
\title{Electron Transport in One-Dimensional Disordered Lattice.}
\author{V. Slavin}
\affiliation{B. Verkin Institute for Low Temperature Physics and 
Engineering of the National Academy of Sciences of Ukraine,
Nauky Ave., 47, Kharkiv, 61103, Ukraine}
\email{slavin@ilt.kharkov.ua}

\author{Y. Savin}
\affiliation{B. Verkin Institute for Low Temperature Physics and 
Engineering of the National Academy of Sciences of Ukraine,
Nauky Ave., 47, Kharkiv, 61103, Ukraine}
\affiliation{Simon Kuznets Kharkov National University of Economics, Nauky Ave., 9a, Kharkiv, 61166, Ukraine}

\author{M. Klimov}
\affiliation{V. N. Karazin Kharkiv National University, 4 Svobody Sq., Kharkiv, 61022, Ukraine}

\author{M. Kiyashko}
\affiliation{Department of Mathematics 432 McAllister Building Penn State University, USA}
\begin{abstract}

We have studied the peculiarities of electron transport in one-dimensional (1D) disordered chain at the presence of correlations between on-site interaction and tunneling integrals.
In the considered models the disorder in host-lattice sites positions is caused by presence of defects, impurities, existence of electron-phonon interaction, e.t.c. 
It is shown, that for certain combination of parameters the localization of electron state, inherited by a various of 1D disordered systems, disappear and electron transport becomes possible.
The parameters of this transport are established.
\end{abstract}
\maketitle

\section{Introduction}
It is well known that in one-dimensional (1D) systems with random potential $U(r)$  (where $r$ is coordinate) the localization of quantum states takes place \cite{And,Lifshits, Phillips1}
for rather wide class of $U$ and $t$. It means, that all one-body wave functions are localized at some finite-size area (this size is called radius of localization $R$) and, hence,
transport in such systems is impossible. For example, saying about electron transport it means that energy-transfer, spin-transfer, information transfer is impossible over 1D  wires with disordered distribution of ions, forming of host-lattice. Note, if we consider finite size chain (with length $N$) and if $R\gtrsim N$ then the effects related to localization may not appear in measurements. As an example, we can cite waveguides, which with high accuracy represent one-dimensional systems with a non-ideal (disordered) surface.

The exact criteria for $U(r)$ and $t(r)$ guaranteeing the localization of quantum states are the following:
Let $U(r)$ and $t(r)$ be random with continuous functions with distribution densities $f$ and $g$:
\begin{enumerate}[label=(\roman*)]
\item space correlations should be absent: ${\rm Cov}[V(r), V(r')]={\rm Cov}[t(r),t(r')]=\delta(r,r'), {\rm Cov}[V(r), t(r')]=0$, where ${\rm Cov}[x,y]=<xy>-<x><y>$. 
\item the distributions should be continuous; $f(x), g(x) >0$ $\forall x\in(-\infty, \infty)$;
\item the distributions should be identical: $f(U(r))=f(U(r'))$,  $g(t(r))=g(t(r'))$ (so called homogeneous in mean). 
\end{enumerate}
Note, that the proofs of corresponding theorems for (i)-(iii) is simple essentially in the case of discrete (lattice) models, where the positions of particles are bonded with host-lattice sites positions $r_n$, $n=1,2,\ldots.$
That is why the main part of the investigations, concerning one-body localization are carried out in the framework of discrete models.

Violation of any of these conditions makes transport possible, but of course does not guarantee its existence. In each case it is necessary to solve the corresponding problem.

For example, in paper \cite{Phillips1,Jitomirskaya1,Jitomirskaya2} in the framework of discrete Schr\"odinger equation  was assumed the existence of correlations between $U$ and $t$ (i.e. requirement  (i) is violated).
As the result, the for certain combination of $U$ and $t$ the localization is absent and electron transport takes place. 

In \cite{Oliveira1,Oliveira2,Sl_Sa} was considered 2-band model (discrete Dirac equation) with dimer correlations of potential $U$ and discrete (Bernoulli) distribution of $f(U)$ (i.e. requirement  (ii) is violated).
As in the case above, for certain combination of $U$ parameters the localization is absent and electron transport takes place. 

The violation of condition (iii) means loss of mean homogeneity. This case is difficult to implement in practice and, hence, rarely studied.

\section{Hamiltonian}
In our study one-body 1D discrete Hamiltonian was chosen in the form:
\begin{equation}
{\hat H} = -\sum_{n} \left(t_n {\hat c}^+_n {\hat c}_{n+1}  + t_{n}^* {\hat c}^+_{n+1} {\hat c}_n\right)+\sum_{n} U_n {\hat c}^+_n {\hat c}_n.
\label{Ham1}
\end{equation}

\noindent Here index $n$ enumerates host-lattice sites; ${\hat c}^+_n$ and ${\hat c}_{n}$ are the creation/annihilation operators of spin-less fermions on site $n$,
$t_{n}$ are the hopping constants and $U_n$ is on-site potential.

The disorder in $U_n$ and $t_n$ is caused by the disorder in host-lattice site positions:  
\begin{equation}
r_n = na_0 + x_n, \quad n=0,1,2,\ldots.
\label{r_n}
\end{equation}

\noindent Here $a_0$ is the distance between the nearest host-lattice sites of ideal (unperturbed) lattice, and   
$x_n$ are random variables, so that: $-c a_0 \leq x_n \leq c a_0$ (in all our calculations $a_0=1$). 
The constant $0\leq c \leq 1/2$ can be considered as a disorder parameter. Such choice of $c$ allows us to avoid overlapping the host-lattice site positions $r_n$ and facilitates the model.

Such disorder can be caused by presence of defects, impurities or phonon. As far as typical phonon frequencies $\omega_{ph}$ are much less than the frequencies of electron jumps $\omega_{el}$ one can consider
the deformations of lattice $x_n$ as {\it static} random variables. 
If so, $U_n$ describes electron-phonon on-site interaction and in the framework of Holstein model \cite{Holstein} it can be written as:

\begin{equation}
U_n \sim  \left({\hat a}^+_n+{\hat a}_n\right) = U_0 + G \left(x_{n+1} - x_{n-1}\right), \quad G, U_0 \in \mathbb{R}, \quad G, U_0 \geq 0, 
\label{U_n}
\end{equation}
\noindent where ${\hat a}^+_n$, ${\hat a}_n$ are the phonon creation/annihilation operators on site $n$ (see e.g. \cite{Phillips1,Gosar,Su}) and $U_0$ and $G$ are some constants.

The tunneling integrals were chosen in rather general form:
$$t_n=t(|r_{n+1}-r_n|)=Ae^{-b |r_{n+1}-r_n|},$$ 
\noindent where $A$ and $b$ are some constants. In the limit of  weak disorder $c \ll 1$ on can expand $t_n$ and in linear approximation we obtain:
$$t_n = V - \Gamma (x_{n+1} -x_n).$$
\noindent where the constants $V, \Gamma \in \mathbb{C}$. Here and further-on we will follow the notation from \cite{Phillips1}. Hence, the tunneling integrals $t_n$ can be written as:

\begin{equation}
t_n =\sqrt{ |V|^2 +  |\Gamma|^2 (x_{n+1} - x_n)^2  -2|V| |\Gamma||(x_{n+1} - x_n)|\cos(\theta)} e^{i\phi_n},
\label{t_n1}
\end{equation}

\begin{equation}
\tan(\phi_n) = {\rm Im}\,\, t_n/ {\rm Re} \,\, t_n.
\label{phi_n}
\end{equation}

As it was shown in Appendix A, the complex phases $\phi_n$ do not affect on our results and can be omitted and instead of (\ref{Ham1}) we will study 
the properties  of real-valued Hamiltonian $\hat{\tilde{H}}$, which is tridiagonal with the following non-zero elements:
\begin{equation}
{\hat {\tilde H}}_{n,n} = U_n, \quad {\hat {\tilde H}}_{n,n-1} = -|t_{n-1}|,  \quad {\hat {\tilde H}}_{n,n+1} = -|t_{n}|
\label{Ham2}
\end{equation}

The proposed model with Hamiltonian (\ref{Ham2}) is similar to those, described in \cite{Phillips1}, but looks more realistic.
We define the disorder in terms of random shifts of host-lattice sites with respect to ``ideal'' positions $na_0$ (see (\ref{r_n})).
The model described in \cite{Phillips1}  defines the disorder in terms of random distances between neighboring host-lattice sites $\Delta_{n,n+1}$.
Besides the differences in distribution functions (distribution of $\Delta_{n,n+1}$ is the distribution of  $x_{n+1} - x_n$),
our model allows us to study consequently the transition from {\it weak} to {\it strong} disorder. 
Actually, increase of $\Gamma$ means increase of fluctuation of host-lattice sites with respect to ``ideal'' positions. 

At the same time, in the model proposed in \cite{Phillips1}, $\Gamma=0$ indeed corresponds to ordered chain, but any small, however nonzero values of $\Gamma$ 
correspond to small mean distances between neighboring host sites. This, on own turn, corresponds to {\it strong} disorder. Moreover, as far as $\Delta_{n,n+1} \geq 0$ $\forall n$
this leads to either an increase, or an decrease of all tunneling integrals $t_n$, depending of the value of parameter $\phi-\theta$ (see, e.g. (2.1) of \cite{Phillips1}).
In our model $t_n$ can be larger or smaller than undisturbed value $t_n=|\Gamma|$ depending on sign of $(x_{n+1} - x_n)$ (see (\ref{t_n1})).

In our model the diagonal terms $U_n\sim \left(x_{n+1} - x_{n-1}\right)$ can be both positive and negative. 
This is physically reasonable: depending on compression or rarefication of host-lattice chain in the vicinity of $n$-th site this term describes either energy gain or energy loss \cite{Holstein}. 
In the model proposed in \cite{Phillips1} $U_n$ are either all negative or all positive.

\section{Results and Discussion}
We studied the solution of the time-dependent Schr\"odinger  equation with Hamiltonian (\ref{Ham2}):
\begin{equation}
{\hat {\tilde H}} |{\tilde \Psi}(x_n, t)\rangle=i\frac{\partial |{\tilde \Psi}(x_n, t)\rangle}{\partial t }
\label{Ham3}
\end{equation}

with the following initial condition:
$$|\Psi'(x_n, t=0)\rangle=\delta(x_n, x_0).$$
This condition corresponds to the localization of an electron at site with index $n=0$ at time moment $t=0$.  

Motion of an electron means ``spreading'' of $|{\tilde \Psi}(x_n, t)\rangle$  as the function of time $t$.  The most informative characteristics describing electron transport in terms of 
$|{\tilde \Psi}(x_n, t)\rangle$ is
time dependence of variance $D(t)$ (see e.g. \cite{Phillips1,Sl_Sa}):

\begin{equation}
D(t)= \langle x(t)^2\rangle - \langle x(t)\rangle^2= m^{(2)}(t) - \left(m^{(1)}(t)\right)^2.
\label{D}
\end{equation}

Here  $m^{(1)}(t)$ and $m^{(2)}(t)$ are the corresponding moments:

\begin{equation}
m^{(k)}(t)=\langle{\tilde \Psi}|(\hat x)^k|{\tilde \Psi}\rangle \,. 
\label{Moments}
\end{equation}

Note that, as it is shown in Appendix \ref{AppendixA}, the moments do not depend on complex phases of off-diagonal matrix elements of the Hamiltonian.
\begin{figure}[ht]
\center{
\includegraphics[width=12.cm]{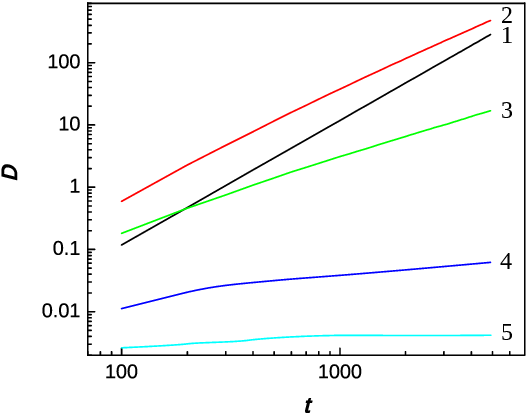} }
\caption{Time dependencies of the variance $D(t)$ (\ref{D}) in double-log scale. In all these calculation $|V|=1$, $G=|\Gamma|$ and the number of host-lattice sites $N=10000$.
Curve 1 (black line) corresponds to the ordered chain ($G=0$), exponent $\alpha=2$ (see (\ref{D_alpha})); this is the case of ballistic transport.
Curve 2 (red line) corresponds to $G=0.25$ and $\theta = 90^0$,  exponent $\alpha=1.6$; this is the case of superdiffusive  transport.
Curve 3 (green line) corresponds to $G=0.2$ and $\theta = 0^0$,  exponent $\alpha=1$; this is the case of diffusive  transport.
Curve 4 (blue line) corresponds to $G=0.5$ and $\theta = 0^0$,  exponent $\alpha=0.35$; this is the case of subdiffusive  transport.
Curve 5 (cyan line) corresponds to $G=0.75$ and $\theta = 0^0$,  exponent $\alpha=0.01$; this is the case of absence of transport.
}
\label{fig1}
\end{figure} 

Approximation of $D(t)$ by power low dependence 
\begin{equation}
D(t)\sim t^\alpha, \quad t \gg 1
\label{D_alpha}
\end{equation} 
\noindent  allows us to establish the existence (or absence) of transport and, if transport exists, the character of the transport.
For example, $\alpha=0$ corresponds to absence of transport, $0 < \alpha < 1$ is so called sub-diffusive transport, $\alpha=1$ is diffusive motion, $1 < \alpha < 2$ is super-diffusive transport and
$\alpha=2$ corresponds to ballistic (free to move) transport.

We studied the solution of (\ref{Ham3}) numerically using 8-th order Runge-Kutta method. To avoid an influence of finite-size effects we checked the probability of finding  an electron on
the last ($N$-th) site of our 1D disordered chain. In all our calculations $|{\tilde \Psi}(x_N, t)|^2 < \varepsilon =10^{-10}$ for all $t$ in the considered time range $0\leq t \leq N/2$.

To check the obtained result we performed additional calculations of (\ref{D}) using spectral data of operator ${\hat {\tilde H}}$ (see (\ref{Ham2})):
the eignevalues $\{\lambda_k\}_{k=1}^N$ and eigenvectors: $\{\psi_k(x_n)\}_{k,n=1}^N$. These values were obtained numerically too.
Such approach is applicable for rather ``small'' systems with length $N\approx 1000-3000$ because calculation time of spectral data increases as  $N^3$.
In terms of spectral data the wavefunction of (\ref{Ham3}) can be written as:

\begin{equation}
|{\tilde \Psi}(x_n,t)\rangle= \sum_{k=1}^N \psi_k(x_n) \psi_k(x_0) e^{i \lambda_k t}.
\label{Psi_Evolution}
\end{equation}

\noindent The vectors $\psi_k(x_n)$ are real-valued and complex conjunction in (\ref{Psi_Evolution}) is omitted. The results obtained using (\ref{Ham3}) and  (\ref{Psi_Evolution}) are in full agreement.

\begin{figure}[ht]
\center{
\includegraphics[width=12.cm]{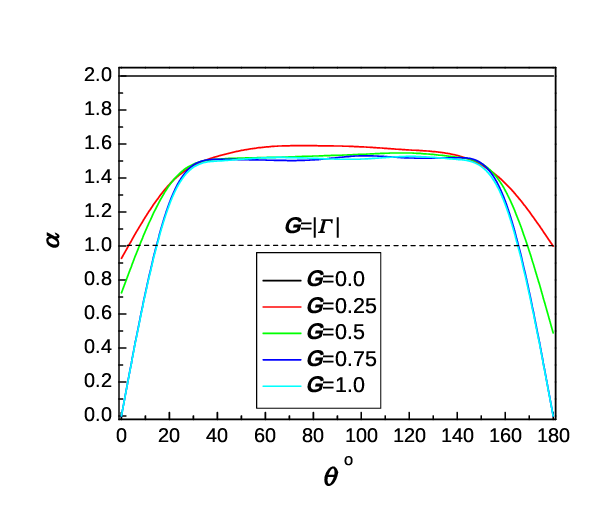} }
\caption{The dependencies of exponent $\alpha$ (see (\ref{D_alpha})) on parameter $\theta$ for different values of $G=\Gamma$. In all these calculations  $V=1$,  and the number of host-lattice sites $N=20000$.}
\label{fig2}
\end{figure}

Typical dependencies $D(t)$ for $N=10000$ and certain combinations of the parameters $G$, $\Gamma$ and $\theta$ are presented in double-log scale in Fig.~\ref{fig1}. In all these calculations $V=1$. 
One can see that all possible transport regimes can be realized in the framework of proposed model: from localization of states up to ballistic transport.

Approximation of $D(t)$ by the expression (\ref{D_alpha}) at $t\gg 1$  allows us to obtain the dependencies of exponent $\alpha$ on tunneling ($V$, $\Gamma$, $\theta$) and hopping ($G$) parameters. 
These dependencies are shown in Fig.~\ref{fig2}. 
One can see that in rather wide range of $0 < G=|\Gamma|\leq 1$ and $20^0 \lesssim \theta \lesssim 160^0$ superdiffusive transport regime is realized. For $0<\theta \lesssim 20^0$ 
and $160^0 \lesssim \theta < 180^0$  transport regime depends on $G, |\Gamma|$: if $0< G=|\Gamma| < 0.22$ the transport is superdiffusive;
if  $G=|\Gamma| > 0.22$ the transport is subdiffusive. 
If $\theta=0^0$ or $\theta=180^0$  and $G=|\Gamma| > 0.5$ electron transport is absent.

It should be noted that the proposed model was formulated within the framework of the linear expansion of on-site potential $U_n$ (\ref{U_n}) and tunneling integrals $t_n$ (\ref{t_n1}) 
with respect to random lattice deformations $x_n$ (\ref{r_n}) i.e. in the limits of  $G\sim |\Gamma| \ll |V|$. Nevertheless, we extended the area of $G$ and $|\Gamma|$ up to $ G, |\Gamma| \leq 1$.
In this case, of course, the Hamiltonian (\ref{Ham2}) and be considered as model only.

\begin{figure}[ht]
\center{
\includegraphics[width=12.cm]{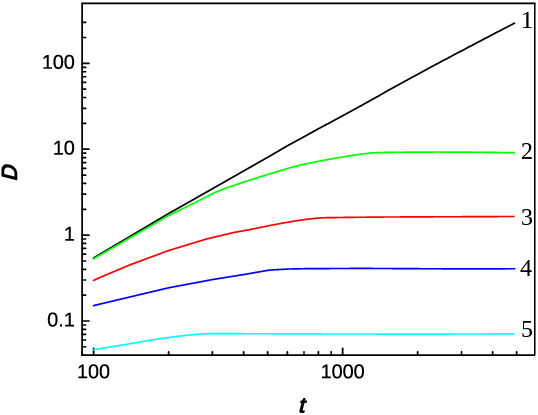} }
\caption{Time dependencies of the variance $D(t)$ (\ref{D}) in double-log scale. In all these calculation $|V|=1$, $\theta=90^0$ and the number of host-lattice sites $N=10000$.
Curve 1 (black line) corresponds to $G=|\Gamma|=0.5$, exponent $\alpha=1.5$ (see (\ref{D_alpha})). 
Curve 2 (red line) corresponds   to $G=0.5$ and $|\Gamma|=0$,  exponent $\alpha=0$. 
Curve 3 (green line) corresponds to $G=0$ and $|\Gamma|=0.5$,  exponent $\alpha=0$. 
Curve 4 (blue line) corresponds  to $G=0.5$ and $|\Gamma|=1$,  exponent $\alpha=0$;
Curve 5 (cyan line) corresponds  to $G=1$ and $|\Gamma|=0.5$,  exponent $\alpha=0$;
}
\label{fig3}
\end{figure}

As was shown in \cite{Phillips1} the case of $G=|\Gamma|$ is of special interest due disappearing  the scattering. Hence, this case is the most prospective for electron transport. 
An influence of $G$ and $\Gamma$ on $D(t)$ is presented in Fig.~\ref{fig3}.

\begin{figure}[ht]
\center{
\includegraphics[width=12.cm]{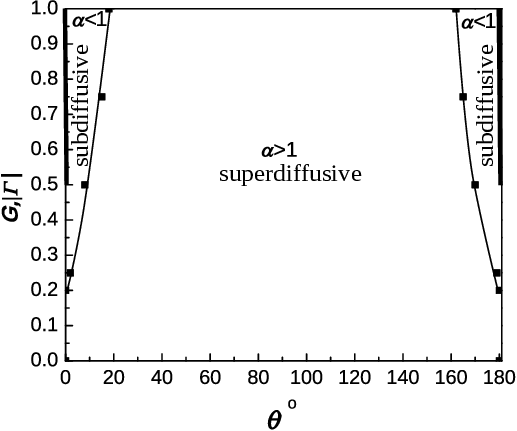} }
\caption{``Phase Diagram'' based on the analysis of the dependence of exponent $\alpha$ (\ref{D_alpha}) on the parameters $G = |\Gamma|$ and $\theta$.
Wide vertical solid lines at $\theta =0^0$ and $\theta=180^0$ correspond to localization area ($\alpha=0$).}
\label{fig4}
\end{figure}

An absence of transport at $\theta=0^0, 180^0$ and $G=|\Gamma|>0.5$ (see Fig~.\ref{fig4}) is of special interest.  To understand the reason of such a peculiarity we performed additional
investigations using Lyapunov exponent \cite{Lifshits,Sl_Sa}. Applying  F\"urstenberg theorem \cite{Furstenberg1,Furstenberg2} one can write Lyapunov exponent $\gamma$  as the following:

\begin{equation}
\gamma(\lambda) = \lim_{n\to \infty} \frac{1}{n} \ln\left(\left\|\prod_{k=n}^1 {\hat T}_k(\lambda)\right\| \right)
\label{gamma}
\end{equation}

Here the transfer matrices ${\hat T}_k$ have the form \cite{Lifshits,Furstenberg1,Sl_Sa}:

\begin{equation}
{\hat T}_k(\lambda)=\left(
\begin{array}{cc} 
\frac{U_k-\lambda}{|t_k|} & -\frac{|t_{k-1}|}{|t_k|}\\ 
1 &0 
\end{array}
\right),
\label{TM}
\end{equation}

\noindent As far as typical values of Lyapunov exponent $\gamma \sim R^{-1} \sim D^{-1/2}(t\to\infty)$, where $R$ is localization radius, the zeros of $\gamma$ indicate on possible delocalization of the states 
(more exactly, the condition $\gamma=0$ is necessary, but not sufficient for delocalization of the states, see e.g. \cite{Lifshits,Sl_Sa}).

\begin{figure}[ht]
\center{
\includegraphics[width=12.cm]{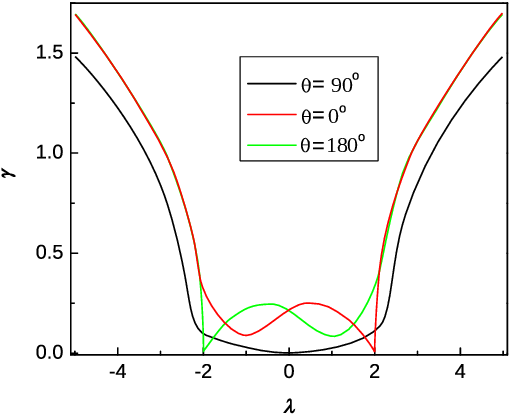} }
\caption{The dependencies of the Lyapunov exponent $\gamma$ on energy $\lambda$. $|V|=1$, $G=|\Gamma|=0.5$ and the number of host-lattice sites $N=10000$.
Black line  corresponds to $\theta = 90^0$. 
Red line corresponds to $\theta = 0^0$ and green line corresponds to  $\theta = 180^0$.}
\label{fig5}
\end{figure}

The results of our calculations are presented in Fig.~\ref{fig5}. 
We see that at $\theta=90^0$ the behavior of Lyapunov exponent $\gamma$ in the neighborhood of root $\gamma(0)=0$ is $\gamma(\lambda)\sim \lambda^\beta$, where $\beta > 1$.
At the same time at $\theta=0^0$ and $\theta=180^0$ the behavior Lyapunov exponent $\gamma$ in the  neighborhood of roots $\gamma(\pm 2)=0$ is $\gamma(\lambda)\sim \left(\pm(2-\lambda)\right)^\beta$, where $\beta < 1$.
It was shown in \cite{Lifshits,Sl_Sa}, that such root-like singularities may not lead to delocalization of states.

\section{Conclusions}

We have studied electron transport properties in 1D disordered chain in the framework of one-body Schr\"odinger equation. 
In the presented model the disorder is caused by an influence of defects, impurities or phonons on the positions of ions, forming host-lattice chain. 
We have shown, that such disorder affects both on the values of on-site potential $U$ and on  tunneling integrals $t$.
Just due to the correlations between random $U$  and $t$ electron transport becomes possible.
We have studied and influence of model parameters on this transport. The areas of localization, superdiffusion and subdiffusion transport have been established.
It has been shown that for certain combinations of model parameters ($|V|=1$, $G=|\Gamma|>0.5$ and for $\theta =0^0, 180^0$) electron transport is impossible,
despite the fact that Lyapunov exponent vanishes in the considered energy region.

\acknowledgements{V. Slavin, M. Klimov and M.  Kiyashko acknowledge the support from the Project IMPRESS-U: N2401227.}

\appendix
\section{}
\label{AppendixA}
Tridiagonal Hermitian matrix $\hat H$ (\ref{Ham1}) can be reduced to symmetric (real-valued) form $\hat{\tilde{ H}}$ using unitary transformation:
\begin{equation}
{\hat {\tilde H}}={\hat U}^{-1}{ \hat H}{\hat U} ,
\label{H_prim}
\end{equation}

\noindent where  unitary  matrix ${\hat U}$ is diagonal:
$$U_{11}=1, \quad U_{nn}= e^{-i\sum\limits_{k=1}^{n-1} \alpha_k}, \quad n=2,3,\ldots.$$
\noindent Here  $\alpha_k$ are the complex phases of off-diagonal elements (\ref{t_n1}), (\ref{phi_n}).
$${\hat {\tilde H}} = {\hat U}^{-1} {\hat H} {\hat U}$$

Let  $|\Psi(x_n, t)\rangle$ be the solution of time dependent Schr\"odinger  equation with Hamiltonian $\hat H$ (\ref{Ham1}) and $|{\tilde \Psi}(x_n, t)\rangle$ be the corresponding solution of  Schr\"odinger  equation with Hamiltonian  $\hat{ \tilde{H}}$ (\ref{Ham2}), (\ref{H_prim}). 

We will show, that the unitary transformation do not affect on the moments $m^{(k)}(t)$ (\ref{Moments}), i.e:

$$m^{(k)}(t)=\langle\Psi|(\hat x)^k|\Psi\rangle=\langle{\tilde \Psi}|(\hat x)^k|{\tilde \Psi}\rangle, \quad  \left((\hat x)^k\right)_{\alpha,\beta} = \alpha^k \delta_{\alpha,\beta} $$

\noindent and, hence, do not affect on $D(t)$ (\ref{D}).

According to the definition 

$$m^{(k)}(t)=\sum_{n} (x_n)^k |\Psi(x_n, t)|^2 $$ 

The relationship between $|\Psi\rangle$ and $|{\tilde \Psi}\rangle$ is:

$$\Psi_j= \sum_{\alpha=1}^N U_{j, \alpha} {\tilde \Psi}_\alpha.$$

\noindent Hence, 

$$m^{(k)}=\sum_{j=1}^N (x^k)_{j,j} \Psi_j \Psi_j^*  =\sum_{j=1}^N \sum_{\alpha=1}^N \sum_{\beta=1}^N (x^k)_{j,j} U_{j, \alpha} U^*_{j, \beta} {\tilde \Psi}_\alpha ({\tilde \Psi}_\beta)^*$$

As far as ${\hat U}$ is unitary, $U^*_{j, \beta}=U^{-1}_{\beta, j}$. Thus,

$$m^{(k)}=\sum_{j=1}^N \sum_{\alpha=1}^N \sum_{\beta=1}^N (x^k)_{j,j} U_{j, \alpha} U^{-1}_{\beta,j} \Psi'_\alpha (\Psi'_\beta)^*=$$

$$ \sum_{j=1}^N \sum_{\alpha=1}^N \sum_{\beta=1}^N  U^{-1}_{\beta,j} (x^k)_{j,j} U_{j, \alpha} \Psi'_\alpha (\Psi'_\beta)^*=$$
$$=\sum_{j=1}^N \sum_{\alpha=1}^N \sum_{\beta=1}^N ({\tilde x}^k)_{\beta,\alpha} \Psi'_\alpha (\Psi'_\beta)^*,$$

\noindent where
$$\left(\hat{\tilde x}^k\right)_{\beta,\alpha}  = \sum_{j=1}^N U^{-1}_{\beta,j} (x^k)_{j,j} U_{j, \alpha}$$
\noindent  are the matrix elements of operator $\hat{x}^k$ in new basis. As far as $\hat{x}^k$ and ${\hat U}$ are diagonal:
$$\left(\hat{\tilde x}^k\right)_{\beta,\alpha}  = \sum_{j=1}^N U^{-1}_{\beta,j} (x^k)_{j,j} U_{j, \alpha} = \sum_{j=1}^N U^{-1}_{\beta,j} \delta_{\beta,j} (x^k)_{j,j} U_{j, \alpha} \delta_{\alpha,j}= 
\left(x^k\right)_{\beta,\alpha} \, .$$


Finally, we have

$$m^{(k)}=\langle\Psi|(\hat x)^k|\Psi\rangle=\sum_{\alpha=1}^N \sum_{\beta=1}^N ({\tilde x}^k)_{\alpha,\beta}  {\tilde \Psi}_\alpha ({\tilde \Psi}_\beta)^*=
\langle {\tilde \Psi}|\hat{x}^k| {\tilde \Psi}\rangle$$

\end{document}